# Modeling and Characterization of Charged Particle Trajectories in an Oscillating Magnetic Field

Dani Irawan[*], Sparisoma Viridi, Siti Nurul Khotimah,
Fourier Dzar Eljabbar Latief, and Novitrian

*Physics Department, Institut Teknologi Bandung, Jalan Ganesha 10, Bandung 40132, Indonesia*
[*]*irawan_dani@yahoo.com*

**Abstract.** A constant magnetic field has frequently been discussed and has been known that it can cause a charged particle to form interesting trajectories such as cycloid and helix in presence of electric field, but a changing magnetic field is rarely discussed. In this work, modeling and characterization of charged particle trajectories in oscillating magnetic field is reported. The modeling is performed using Euler method with speed corrector. The result shows that there are two types of trajectory patterns that will recur for every $180nT_0$ ($n$ = 0, 1, 2, ..) in increasing of magnetic field oscillation period, where $T_0$ is about $6.25 \times 10^{-7}$ s.

## INTRODUCTION

Many books have shown how a constant magnetic field affects a charged particle motion through the Lorentz force equation. Some of the trajectory patterns are interesting, e.g. the cycloid and the helix as they are shown in Figure 1 and 2, respectively [1].

Different than motion within a constant magnetic field, motion within the influence of a non-constant magnetic field is rarely discussed. We tried to model the non-constant case using Euler method with speed corrector [2] in absence of electric field.

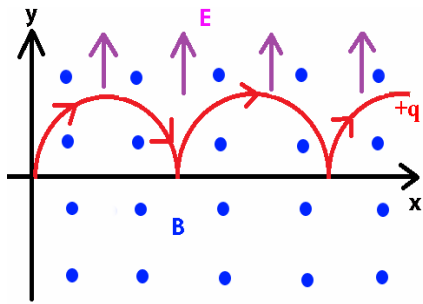

**FIGURE 1.** Cycloid motion is caused by constant magnetic field and constant electric field that are perpendicular.

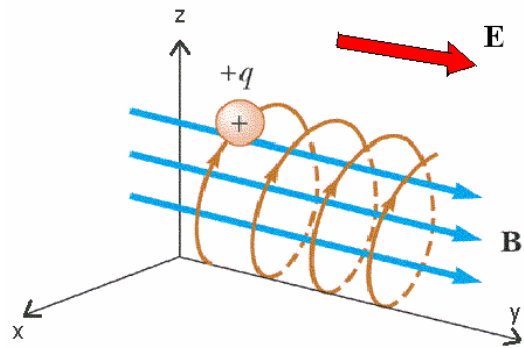

**FIGURE 2.** Helix motion is produced if constant electric field and constant magnetic field are parallel.

Those two example show that there are some interesting patterns that can be observed in charged particle motion by varying some parameters, e.g. direction of magnetic field to electric field.

The reason why we need the speed corrector is to ensure that change of particle kinetic energy is only due to change of magnetic field and not other factors.

## SPEED CORRECTOR

Suppose that our particle initial velocity is as follow

$$v_{initial} = v_x \hat{i} + v_y \hat{j} \qquad (1)$$

Thus, its initial speed is

$$|v_{initial}| = \sqrt{v_x^2 + v_y^2} \qquad (2)$$

After a time step $\Delta t$, its velocity changes due to the presence of constant perpendicular magnetic field [2]

$$v_{new} = (v_x + \frac{qv_y B_z}{m}\Delta t)\hat{i} + (v_y - \frac{qv_x B_z}{m}\Delta t)\hat{j}, \qquad (3)$$

or in term of speed

$$|v_{new}| = \sqrt{(v_x^2 + v_y^2)\cdot\left(1+\left(\frac{qB_z}{m}\Delta t\right)^2\right)}. \qquad (4)$$

To correct the new speed, we define a speed corrector

$$\alpha = \frac{|\vec{v}_{initial}|}{|\vec{v}_{new}|} = \frac{1}{\sqrt{1+\left(\frac{qB_z}{m}\Delta t\right)^2}}, \qquad (5)$$

then we compute the corrected velocity as

$$\vec{v}_{corrected} = \vec{v}_{new} \cdot \alpha . \qquad (6)$$

Equation (6) will ensure that the speed is constant if the magnetic field is constant, but the velocity can still keep on changing. Using Euler method without any corrector will not produce a constant speed for this case [2].

## CHARGED-PARTICLE MODELING

In this model, it is assumed that there is only one external force affecting the particle. This external force is the magnetic force which follows [1]

$$\vec{F}_B = q\vec{v} \times \vec{B}, \qquad (7)$$

where particle velocity $\vec{v}$ and the magnetic field $\vec{B}$ are chosen to be perpendicular.

From Equation (7), using Newton's second law of motion we can get the acceleration of the particle for every time $t$ [3, 4]. Then, using the acquired acceleration, we can calculate the velocity and position using Euler method [5]. To ensure that particle kinetic energy is constant for constant magnetic field case, the velocity is always corrected using Equation (6) before it is used to calculate the new position. For non-constant magnetic field case we can only assume that the change of particle kinetic energy only due to the change of magnetic field since the speed corrector is already applied.

The characterization of trajectory patterns are done by changing the initial condition of the particle or changing the magnetic field, i.e. the particle initial speed or direction, or the magnetic field oscillation period.

The magnetic field used in the model is defined as follows

$$B(t) = B_0 \sin\left(\frac{2\pi}{T}t\right), \qquad (8)$$

with $B_0$ is the maximum magnetic field strength and $T$ is the period of the magnetic field oscillation. We can change all these parameters to observe how it will affect the particle motion.

## RESULTS

Two different cases are calculated in this work. In the first case constant magnetic field is used, while oscillating magnetic field is used in the second.

### Constant Magnetic Field Case

Before calculating particle motion in an oscillating magnetic field, particle trajectory in a constant magnetic field should be first checked, whether it forms a circular path or not.

For this constant magnetic field case following parameters are used
1. initial position at (0, 0),
2. initial velocity is 500 m/s parallel to the positive-*x* axis.

Some results are shown in Figure 3 and 4. It can be seen from those figures that the trajectories have a circular path whose radius depends on chosen time step $\Delta t$. It also has been observed that radius error is dependent on product of $B\Delta t$ as shown in Figure 5.

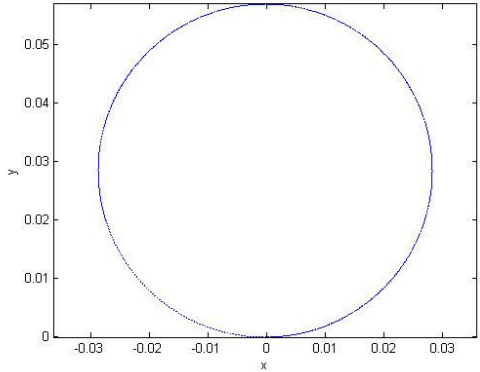

**FIGURE 3.** Particle trajectory for constant magnetic field with $B = 10^{-7}$ T and $\Delta t = 10^{-6}$ s.

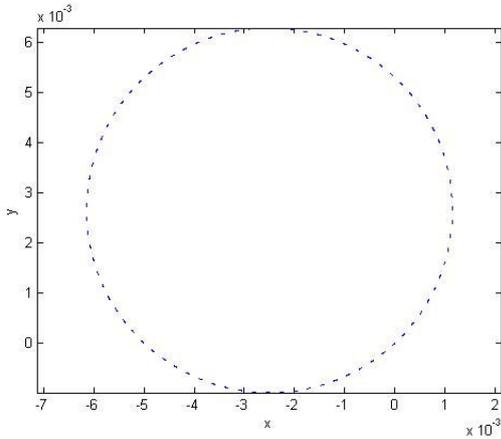

**FIGURE 4.** Particle trajectory for constant magnetic field with $B = 10^{-5}$ T and $\Delta t = 10^{-5}$ s.

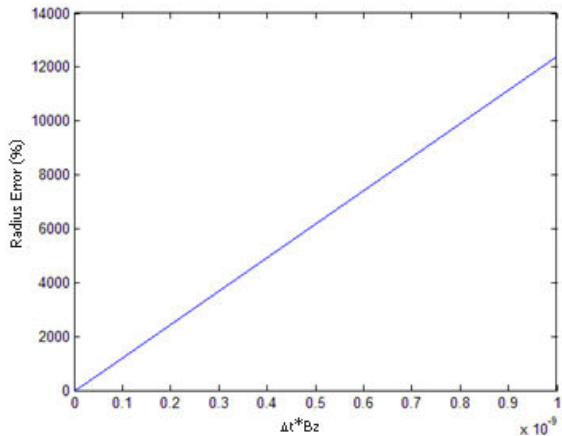

**FIGURE 5.** Radius error dependence on product of $B\Delta t$.

Figure 5 shows that for getting smaller error the product of $B\Delta t$ should be less than $10^{-9}$.

## Oscillating Magnetic Field Case

For this non-constant magnetic field case following parameters are used
1. initial position at (0, 0),
2. the magnetic field used in this case follows Equation (8),
3. $B_0$ is $10^{-6}$ T and $\Delta t$ is $10^{-7}$ s, this $B\Delta t$ produces radius error about 0.115%.

By changing the oscillation period of the magnetic field, we found two different types of trajectory pattern. The first pattern is a flowing pattern as shown in Figure 3, and the second pattern is a static pattern shown in Figure 4. In static pattern the particle is localized in some region, while it is not in flowing pattern.

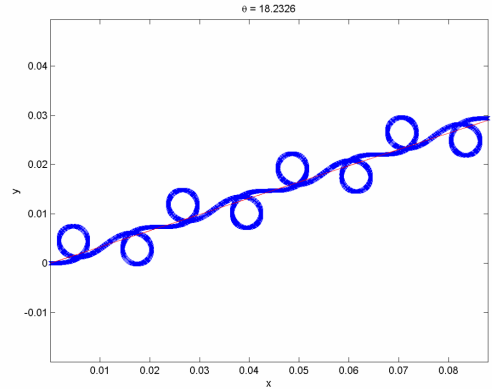

**FIGURE 6.** A typical flowing trajectory pattern.

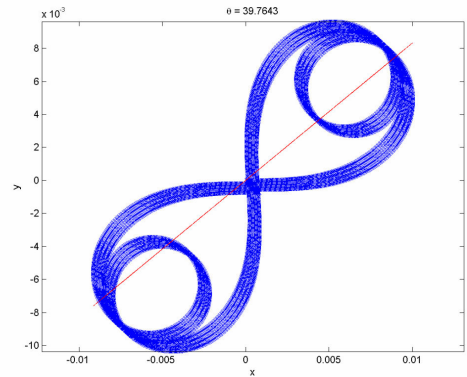

**FIGURE 7.** A typical static trajectory pattern.

The second pattern shows a possibility to make electron trap using oscillating magnetic field.

## Particle Initial Direction

To observe the influence of particle initial direction, all parameter except the initial direction is set. The following result is done by setting the magnetic field oscillation period at 400 kHz and speed at 500 m/s.

The particle path orientation is the angle between curve-fitting line of the particle trajectory and positive-x axis and it is given the symbol $\theta$. The particle initial direction is then given the symbol $\theta_0$. The result shows that $\theta$ value will change exactly the same as $\theta_0$. This result can be seen in Figures 8.

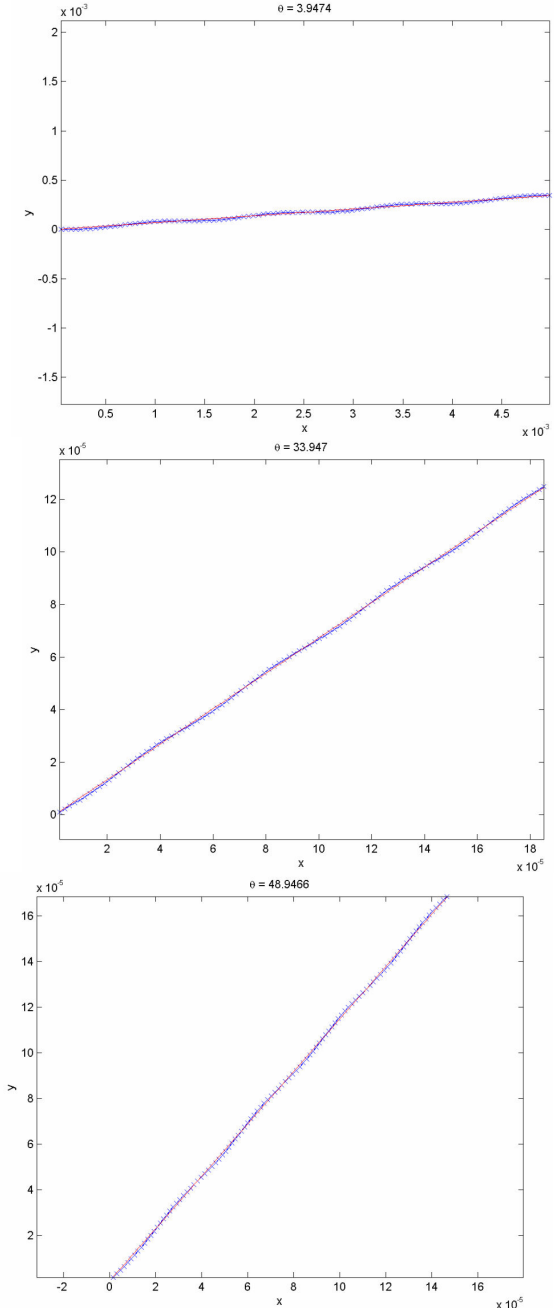

**FIGURE 8.** Fitting line of particle trajectory for initial direction $\theta_0$ : 0 ° (top), 30 ° (middle) and 45 ° (bottom).

## Magnetic Field Oscillation Period

The model shows that magnetic field oscillation period can change the particle orientation. Because the initial direction can affect orientation, we present the effect of magnetic field oscillation period as orientation change $(\theta - \theta_0)$ and given it the symbol $\Delta\theta$. The $\Delta\theta$ for some oscillation period are presented in Figure 9 while their detail trajectories are presented in Figure 10.

The area marked in Figure 9 is the critical area. In this area, the type of trajectory patterns is changing from flowing pattern to static pattern and the orientation cannot be determined anymore.

The $\Delta\theta$ will change 1 ° for every 6.25 × 10⁻⁷ s (we call this value as $T_0$) change in oscillation period. By changing the x-axis of Figure 9 to $T/T_0$ (Figure 13) we can see that $\Delta\theta$ is periodic and will recur for every 180 $T/T_0$.

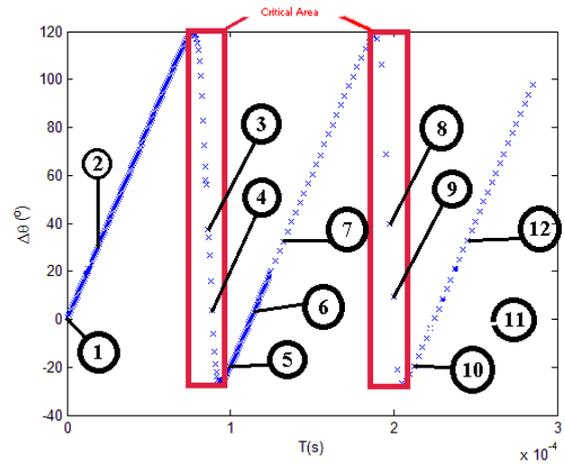

**FIGURE 9.** $\Delta\theta$ for different oscillation period of magnetic field.

## Particle Speed

Particle speed does not show any influence on fitting line orientation and type of trajectory pattern. Particle speed only changes the path length (See Figure 12 and 13). The ratio of values in vertical and horizontal axis remains the same even the particle speed is changed. It means that the particle path orientation does not change.

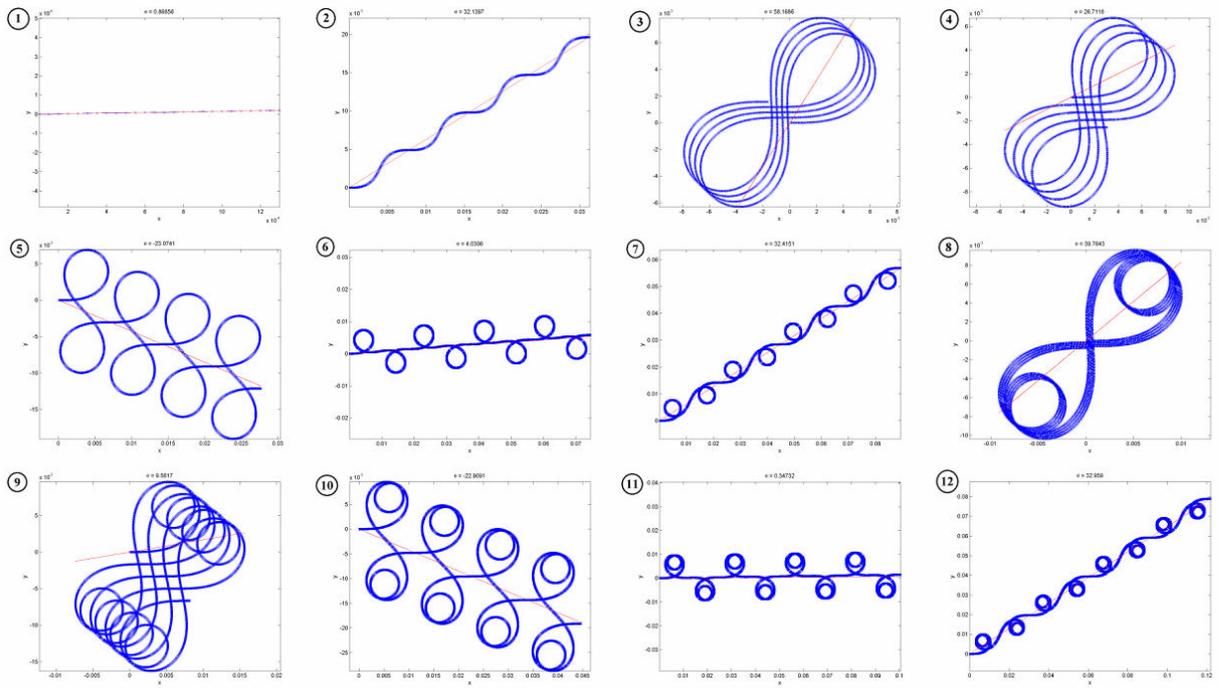

**FIGURE 10.** Particle trajectory patterns for different oscillation period of magnetic field.

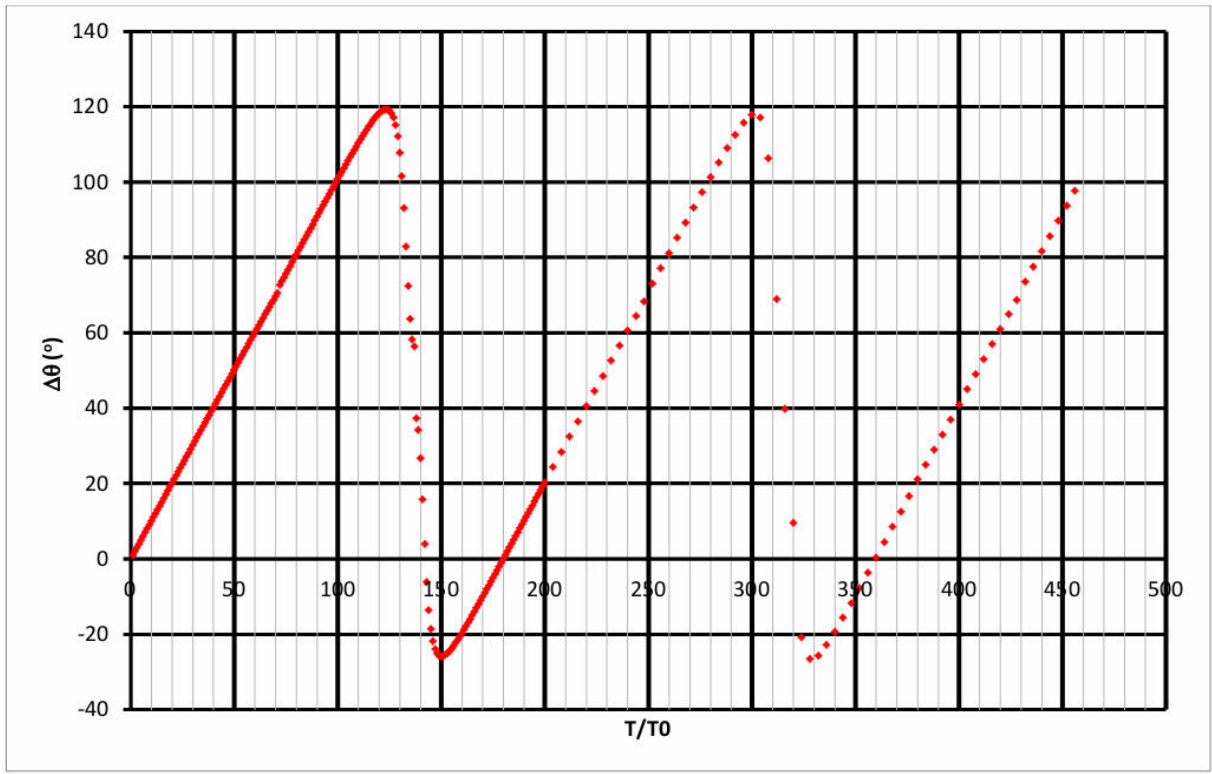

**FIGURE 11.** $\Delta\theta$ for different $T/T_0$.

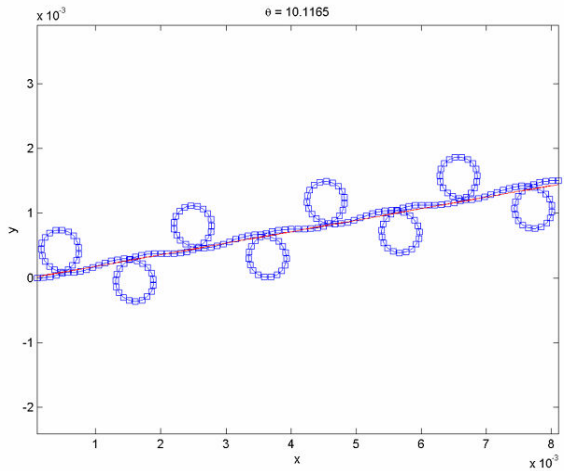

**FIGURE 12.** Particle trajectory using $1.1875 \times 10^{-4}$ s magnetic field oscillation period and 50 m/s speed.

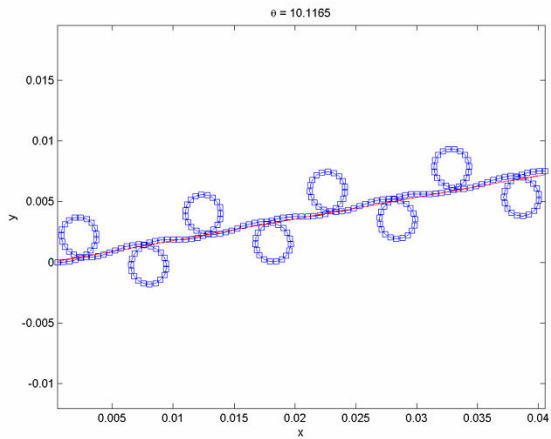

**FIGURE 13.** Particle trajectory using $1.1875 \times 10^{-4}$ s magnetic field oscillation period and 250 m/s speed

## CONCLUSION

It can be concluded that there are two types of trajectory pattern: flowing and static. Particle path orientation depends on magnetic field oscillation period. It is increased by 1 ° for every $T_0 = 6.25 \times 10^{-7}$ s in increasing of magnetic oscillation period except in the critical area and will recur for every 180 $T/T_0$.